\documentclass[amsmath,amssymb,aps,prl,superscriptaddress,twocolumn,floatfix]{revtex4-2}
\usepackage{graphicx}
\usepackage{dcolumn}
\usepackage{bm}
\usepackage[usenames]{color}
\usepackage{comment}
\usepackage{wasysym}
\usepackage{hyperref} 
\usepackage{physics}
\usepackage{bookmark}
\usepackage{mathtools}
\usepackage{amsmath}
\begin{document}

\title{Stable magic angle in twisted Kane-Mele materials}

\author{Cheng Xu}
\email{chengxu.physics@gmail.com}
\affiliation{Department of Physics and Astronomy, University of Tennessee, Knoxville, Tennessee 37996, USA}
\affiliation{Department of Physics, Tsinghua Univeristy, Beijing, China}

\author{Yong Xu}
\affiliation{Department of Physics, Tsinghua Univeristy, Beijing, China}

\author{Wenhui Duan}
\affiliation{Department of Physics, Tsinghua Univeristy, Beijing, China}

\author{Yang Zhang}
\email{yangzhang@utk.edu}
\affiliation{Department of Physics and Astronomy, University of Tennessee, Knoxville, Tennessee 37996, USA}
\affiliation{Min H. Kao Department of Electrical Engineering and Computer Science, University of Tennessee, Knoxville, Tennessee 37996, USA}


\begin{abstract}

We propose that flat bands and van Hove singularities near the magic angle can be stabilized against angle disorder in the twisted Kane-Mele model. With continuum model and maximally localized Wannier function approaches, we identify a quadratic dispersion relationship between the bandwidth, interaction parameters versus the twist angle, in contrast to twisted bilayer graphene (TBG). Introducing Kane-Mele spin-orbit coupling to TBG greatly reduces the fractional Chern insulator indicator and enhances the stability of fractional Chern states near the magic angle, as confirmed by exact diagonalization calculations. Moreover, in twisted bilayer Pt$_2$HgSe$_3$ with intrinsic Kane-Mele spin-orbit coupling, we identify a topological flat band at a large twist angle around 4 degrees.

\end{abstract}

\maketitle

{\it Introduction.---}
Moir\'e materials \cite{kennes2021moire} have opened a new venue in correlation physics. Moir\'e patterns are long-wavelength interference patterns resulting from lattice mismatch in layered materials. The weak interlayer coupling in two-dimensional van der Waals materials provides a highly tunable platform for studying strongly correlated electrons through variations in twist angle, gate voltage, pressure, and strain. In the case of graphene with Dirac band, magic-angle twisted bilayer graphene has demonstrated a wide range of phenomena such as Van Hove singularities \cite{li2010observation}, superconductivity \cite{cao2018b,yankowitz2019tuning,oh2021evidence,saito2020independent}, correlated insulating states \cite{Cao2018c,Jiang2019b}, quantum anomalous Hall effect \cite{Serlin2020,nuckolls2020strongly}, and fractional Chern insulator \cite{xie2021fractional}. 

The central challenge in the experimental study of twisted bilayer graphene is the ultra-sensitive nature of the flat bands, which appear only at the magic angle. According to the original Bistritzer-MacDonald work \cite{bistritzer_moire_2011}, the Dirac velocity (or bandwidth) of moir\'e band exhibits a linear dependence on the twist angle. However, in realistic devices, angle disorder is inevitable due to lattice relaxation, strain and specific fabrication methods. Even a small amount of angle disorder can significantly increase the Dirac velocity and moir\'e bandwidth, thereby destroying correlated Chern states and superconductivity \cite{lau2022reproducibility}.

In this work, we aim to provide the guiding principles for stabilizing strongly correlated phases (especially the integer and fractional Chern states) in twisted bilayer graphene. Previous experiments suggest that the correlated states in magic-angle twisted bilayer graphene (MATBG) can be stabilized by adjacent transition metal dichalcogenides (TMDs) \cite{polski2022hierarchy}. We find that encapsulating MATBG with TMD monolayers\cite{avsar2014spin,island_spinorbit-driven_2019,tan2024topological,bhowmik2023spin} enhances its electronic stability against angle disorder due to the proximitized spin-orbit coupling (SOC). Using a combined continuum model and tight-binding approach, we identify a quadratic dispersion relationship between bandwidth and twist angle, with the magic angle remaining unchanged under proximitized SOC. Additionally, the density of state peaks \cite{li2010observation} around the magic angle are also stabilized, exhibiting a negligible 8\% reduction in variation due to angle disorder. Beyond the single particle band dispersion, the indicators of fractional Chern insulator (FCI) \cite{claassen_position-momentum_2015, roy_band_2014, ledwith_fractional_2020-1} are significantly reduced compared to the original TBG model. Exact diagonalization calculations further reveal a wide region of stabilized FCI states around the magic angle.

Besides graphene systems, the two-dimensional quantum spin Hall insulator Pt$_2$HgSe$_3$ \cite{vymazalova2012jacutingaite,marrazzo2018prediction,facio2019dual}, characterized by the Kane-Mele model with a significantly larger spin-orbit coupling strength, emerges as a promising candidate. We calculate the magic angle, van Hove singularity and FCI indicator in twisted Pt$_2$HgSe$_3$, and suggest it as a robust candidate system for realizing zero-field fractional Chern insulators other than present semiconductor and pentalayer graphene moir\'e materials \cite{park2023observation,xu2023observation,lu2024fractional}.

 \begin{figure*}
  \includegraphics[width=\hsize]{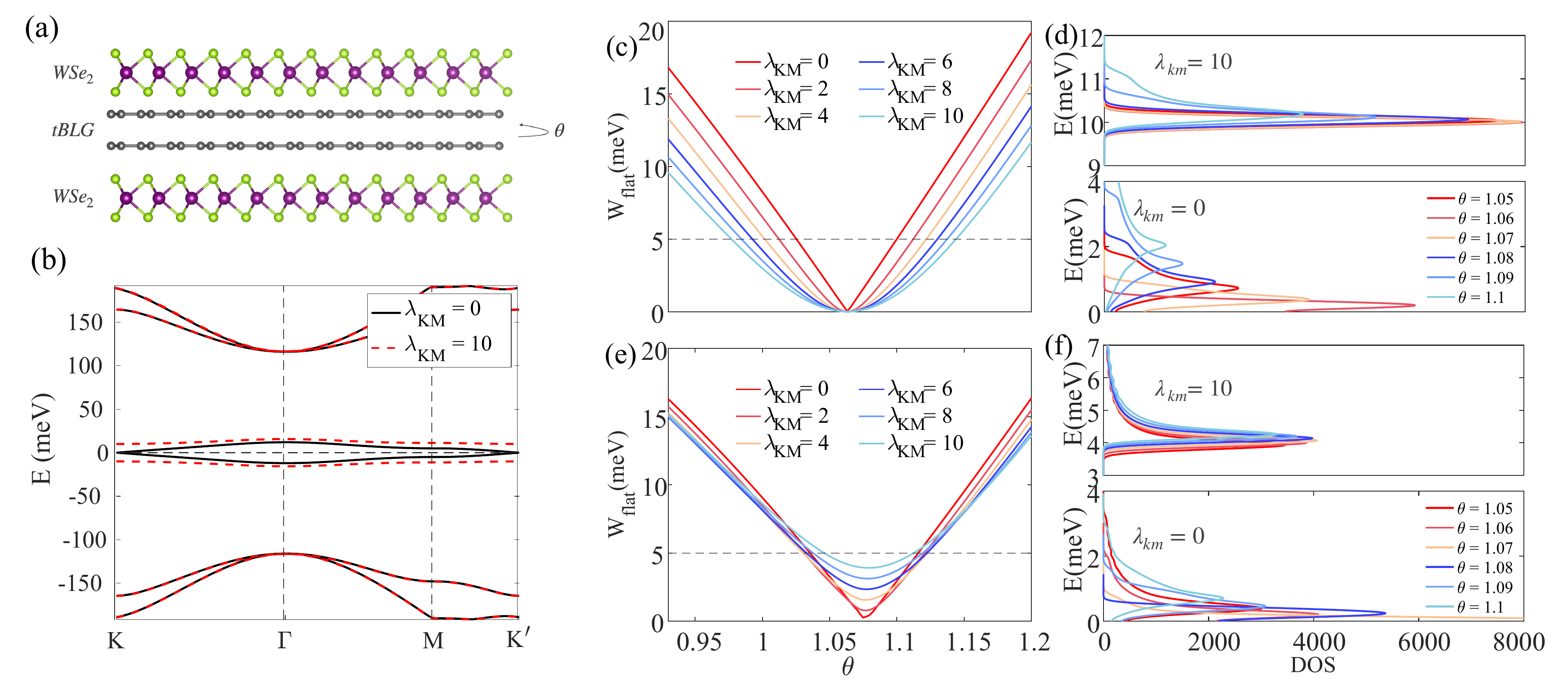}
  \caption{(a) The proposed experimental setup, which induces Kane-Mele SOC in twisted bilayer graphene. (b) The bands of TBG with ($10$ meV)/without Kane-Mele SOC in the chiral limit at 1.15$^\circ$, which is calculated with continuum model.
  (c) The relation between the bandwidth and twisted angle with varing Kane-Mele SOC in the chiral limit. (d)(up) The DOS of the lowest conduction band at different twist angle with $\lambda_{KM}=10$ meV. (d)(down) The same as up one but without SOC. (e)/(f): The same as (c)/(d) but with finite AA hopping $w_0 = 0.7 w_1$. }
  \label{Fig1}
\end{figure*}

{\it The twisted Kane-Mele model---}
Expanding beyond TBG continuum model $H_{BM}$, the twisted Kane-Mele model incorporates an additional Kane-Mele SOC term as follows:
\begin{equation}
	\begin{aligned}
		 	H=H_{BM}\otimes s_0+\lambda_{KM}\sigma_z s_z\tau_0\\
		 	H_{BM}=\begin{bmatrix}
		\hbar v_F\boldsymbol{\hat{k}\cdot \sigma} & \boldsymbol{\hat{T}(r)}\\
		 \boldsymbol{\hat{T}^{\dagger}(r)} & \hbar v_F\boldsymbol{\hat{k}\cdot \sigma}
	\end{bmatrix}
	\end{aligned}
\end{equation}
where the Pauli matrix $\boldsymbol{\sigma}$ denote the pseudo spin of the sub-lattice A and B, $\boldsymbol{s}$ is the Pauli matrix that denote the spin, $\lambda_{KM}$ stands for the strength of Kane-Mele SOC,  $\tau$ is the Pauli matrix corresponds to the pseudo spin of layer and the $v_F$ stands for the Fermi velocity of the single layer and $H_{BM}$ denotes the Hamiltonian of BM model \cite{bistritzer_moire_2011}. The off-diagonal term $\boldsymbol{\hat{T}(r)}=\sum_{i=1}^{3}T_i\mathrm{e}^{-i\boldsymbol{q_i\cdot r}}
$ is the interlayer coupling and 
\begin{equation}
	T_i=w_0\sigma_0+w_1\bigg[\sigma_x\cos(i-1)\phi+\sigma_y\sin(i-1)\phi\bigg]
\end{equation}
is the hopping matrix related to the three different hopping directions $\boldsymbol{q_1,q_2,q_3}$ in momentum space. $w_0$ denote the hopping amplitude between the same sub-lattice  and $w_1$ stands for the hopping between different sub-lattice. The coupling momentum is given by:
  $\boldsymbol{q_1}=k_{\theta}(0,-1)$,$\boldsymbol{q_2}=k_{\theta}(\frac{\sqrt{3}}{2},\frac{1}{2})$ and $\boldsymbol{q_3}=k_{\theta}(\frac{\sqrt{3}}{2},-\frac{1}{2})$, where $k_{\theta}=2|\boldsymbol{K}|\sin\frac{\theta}{2}$ and $|\boldsymbol{K}|=\frac{4\pi}{3a}$ is the Dirac momentum of graphene and $a$ stands for the crystal constant.

Recent experiments and density functional theory (DFT) studies \cite{li_twist-angle_2019,alsharari2016mass,yang2018effect,avsar2014spin,island_spinorbit-driven_2019} have shown that SOC in graphene can be enhanced through proximity effects with transition metal dichalcogenide materials. 
The proximity effects can induce a combination of Ising, Rashba, and Kane-Mele SOC, depending on the actual device geometries. When bilayer graphene is encapsulated by TMDs both above and below, as illustrated in Fig. \ref{Fig1}(a), experiments \cite{wakamura_strong_2018,wakamura_spin-orbit_2019} have found that Kane-Mele SOC plays a dominant role in spin relaxation near the Dirac point. Given that electrons near the Dirac point form the building blocks of moiré bands, we propose that the twisted Kane-Mele model effectively describes the low-energy physics in MATBG with two adjacent WSe$_2$ monolayers.

{\it Band width and stable van Hove singularity.---} 
As the first step, we illustrate how the band structure is deformed by Kane-Mele SOC using the pseudo Landau level representation \cite{liu_pseudo-landau-level_2019,tarnopolsky_origin_2019} of the flat band. Due to the $C_{2z}\mathcal{T}$ symmetry, the spin-up and spin-down bands share the same band structure, allowing us to focus solely on the spin-up bands for simplicity. In the chiral limit ($w_0 = 0$), the Hamiltonian for the flat band can be approximated as:
\begin{equation}
	H_{KM}=\lambda_{KM }\begin{bmatrix}
		1 & 0\\
		0 & -1
	\end{bmatrix}
\end{equation}
with the basis: $	\Psi=[\psi_{A1\uparrow}\  \ \psi_{B2\uparrow}\ ]^{T}$, where 
$\psi_{A1\uparrow} \ , \psi_{B2\downarrow}$ stand for the lowest Landau Level; A, B denote the sub-lattice, the index $1$ and $2$ stand for the combined layer basis. It is important to note that this Hamiltonian is a diagonal matrix, meaning the wave function of the flat band (analogous to the lowest Landau level here) remains unchanged by the Kane-Mele SOC, resulting only in an energy shift. This observation aligns with our numerical results shown in Fig. \ref{Fig1}(b). However, away from the magic angle, the pseudo-Landau level picture becomes inexact, leading to a finite bandwidth. In experiments, due to lattice relaxation effects \cite{nam2017lattice,cantele2020structural,xie2023lattice}, the ratio $w_0/w_1 \approx 0.7 \sim 0.8$, indicating that the chiral limit is not an accurate approximation. When AA hopping is considered, the lowest Landau level couples with higher pseudo-Landau levels in a different subspace with the opposite pseudo-magnetic field, resulting in a finite bandwidth.

As we demonstrate below, the twist angle dependent bandwidth is significantly altered by the Kane-Mele SOC, potentially aiding the correlated state in overcoming angle disorder in experiments. In the chiral limit, this relationship can be derived using a simplified plane wave basis \cite{bernevig2021twisted,bistritzer_moire_2011} (detailed in the supplementary material), and is given by:
  \begin{equation}
 	W=\sqrt{(\hbar v_F|\boldsymbol{K}|\theta-2w_1)^2+\lambda_{KM}^{2}}-\lambda_{KM}
 \end{equation}
 
When $\lambda_{KM}=0$, the dispersion relation reads: $W=|\hbar v_F|\boldsymbol{K}|\theta-2w_1|$, which is a linear function around the magic angle ($\theta_{m}= \frac{2w_1}{\hbar v_F|\boldsymbol{K}|}$). The large Fermi velocity of graphene results in a steep slope of this linear function, meaning that even a slight change in the twist angle can cause a dramatic increase in the bandwidth. This makes it challenging to observe flat band-related phenomena due to the unavoidable angle disorder in experimental devices. Fortunately, we find that the inclusion of Kane-Mele SOC can help overcome this obstacle, as shown below. With the addition of Kane-Mele SOC, the linear dispersion relation transitions into a quadratic dispersion relation:
\begin{equation}
\begin{aligned}
\frac{	\partial W}{\partial \theta}\bigg|_{\theta= \theta_{m}}&=0\\
	\frac{	\partial^2 W}{\partial \theta^2}\bigg|_{\theta= \theta_{m}}  &=\frac{\hbar^2|\boldsymbol{K}|^2v_F^2}{\lambda_{KM}}
\end{aligned}
\end{equation}
The quadratic dispersion relation around the magic angle renders the bandwidth less sensitive to variations in the twist angle, thereby stabilizing the electronic structure against angle disorder in the device. For example, a mere 0.01$^\circ$ change in the twist angle can cause a noticeable shift in the position of the van Hove singularity, as shown in the bottom panel of Fig. \ref{Fig1}(d). Conversely, when the Kane-Mele SOC is included, the position of the van Hove singularity remains almost unchanged. Additionally, it is crucial to note that the second derivative of the dispersion relation is inversely proportional to the strength of the Kane-Mele SOC. This suggests the potential benefits of identifying materials capable of inducing stronger proximitized Kane-Mele SOC, which would result in a more stable electronic structure.

All the analyses presented above are based on the analytic expressions derived in the chiral limit. However, our conclusions remain valid even when deviating from the chiral limit, as demonstrated in Fig. \ref{Fig1}(e-f). When AA hopping is included, a comparison between Fig. \ref{Fig1}(e) and Fig. \ref{Fig1}(c) shows that the quadratic dispersion is preserved. Although the minimal bandwidth cannot reach zero due to the strong coupling between the lowest and higher pseudo-Landau levels, it remains small compared to the interaction scale around the magic angle. Additionally, as shown in Fig. \ref{Fig1}(d), the van Hove singularity is also stabilized, with its position remaining nearly invariant in the top panel, while it shows significant changes with the twist angle in the bottom panel. In summary, the inclusion of Kane-Mele spin-orbit coupling stabilizes the electronic structures near magic angle, which is beneficial for investigating flat band phenomena experimentally, even in the presence of unavoidable angle disorder in real devices.


{\it The interaction parameters derived from real space Wannier functions.---} 
 \begin{figure}
  \includegraphics[width=1.0\hsize]{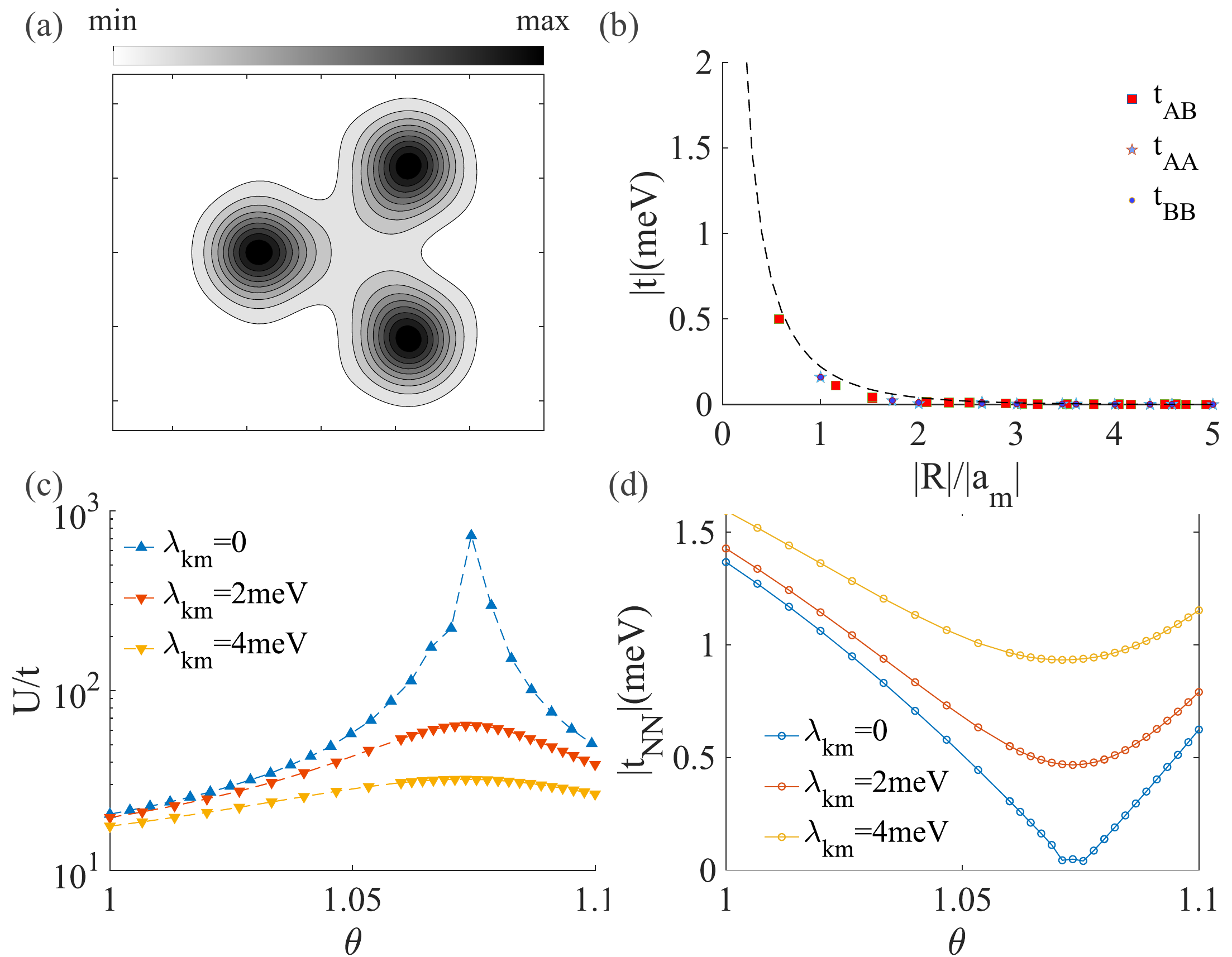}
  \caption{(a) The charge distribution of the maximally localized Wannier functions at 1.08$^\circ$. (b) The tight-binding parameter (1.08$^\circ$) from the MLWF-based tight-binding model, where $t_{AB}$ represents the hopping amplitude between Wannier functions centered on the AB (Wannier A) and BA (Wannier B) sites. The hopping amplitude decreases rapidly with distance. (c) The relation between $U/t$ and twist angle, where $U$ is the onsite repulsion from the MLWF and $t$ is nearest-neighbor hopping amplitude. (d) The relation between nearest-neighbor hopping and the twist angle at different strength of Kane-Mele SOC. We note that $w_0 = 0.7 w_1$ is used in (a)-(d).}
  \label{Fig2}
\end{figure}
In the previous section, we analyzed how bandwidth and density of states (DOS) vary with the twist angle under the influence of Kane-Mele SOC. In this section, we shift to real space and construct maximally localized Wannier functions (MLWFs) for twisted bilayer graphene with Kane-Mele SOC. This approach allows us to evaluate the impact on the interaction parameter in real space. Following the methods outlined in \cite{koshino2018maximally, kang2018symmetry, xu2024maximally}, we obtain Wannier functions with three peaks localized at the AA regions, though they are centered on the AB/BA regions. As shown in Fig. \ref{Fig2}(b), the tight-binding model is short-range and the Wannier functions are well-localized.

Fig. \ref{Fig2}(d) confirms that large Kane-Mele SOC results in a flatter relationship between single-particle hopping and twist angle. We then compute the Hubbard $U$ from the Wannier functions and plot $U/t$ versus $\theta$ in Fig. \ref{Fig2}(c). Our results indicate that the difference of Hubbard model parameters between the magic angle and other twist angles is reduced, suggesting that the relative interaction strength becomes less sensitive to angle disorder. This reduced sensitivity in the electronic structure to twist angle may enhance the reproducibility in the fabrication and study of moiré materials \cite{lau2022reproducibility}.

{\it Fractional Chern insulator.---} 
We have demonstrated that Kane-Mele SOC significantly stabilizes the electronic structure of twisted bilayer graphene, as evidenced by improvements in bandwidth, van Hove singularity, and interaction parameters. This stabilization effect applies generally to correlated states, though we focus specifically on the fractional Chern insulator in this context. While bandwidth and van Hove singularity describe aspects of band dispersion, they do not fully account for electron dynamics; band geometry is also crucial. To characterize the stability of FCI states, we analyze fluctuations in Berry curvature ($\sigma_F$) and deviations from the trace condition (T), as detailed in the supplementary material.

\begin{figure*}
\includegraphics[width=\hsize]{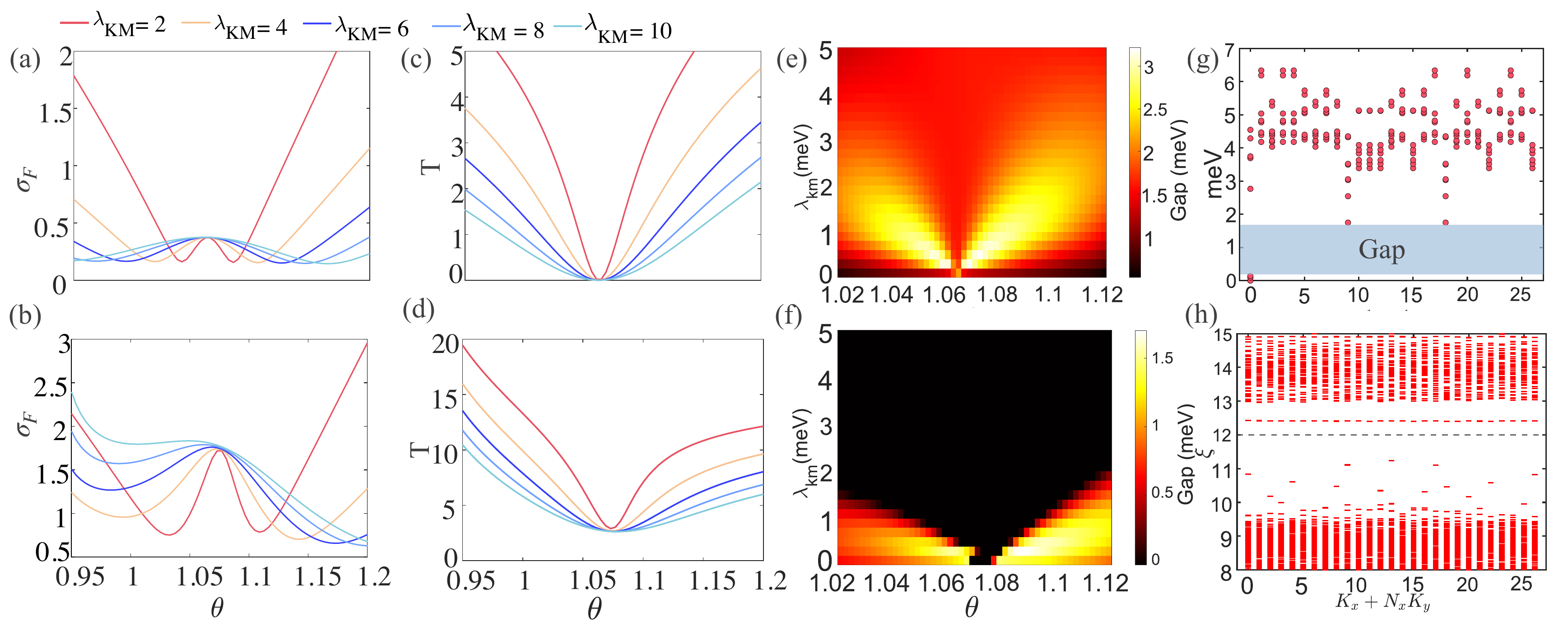}
     \caption{
      (a) The Berry deviation ($\sigma_F$) as a function of twist angle with series value of Kane-Mele SOC in the chiral limit. (b) The same as (a) but with $w_0=0.7w_1$. (c) The relation between the Trace condition(T) and twist angle at a series of strength of Kane-Mele SOC in the chiral limit. (d) The same as (c) but with $w_0=0.7w_1$. (e)/(f) The gap of FCI states as a function of twisted angle and Kane-Mele SOC with $w_0=0$/$w_0=0.7w_1$. (g) The many body spectrum of FCI in chiral limit at $\theta=1.06^\circ,\lambda_{KM}=2 meV$, the gap is denoted by the blue region. (h) The particle entanglement spectrum calculated on the 27 sites cluster with cutting corresponding to $N_A=3$, and there are 1710 states below the gap which matches the counting of quasi-hole excitation of fractional quantum Hall effect.}
     \label{Fig3}
\end{figure*}

To favor a fractional Chern insulator, the band geometry is suggested to minimize both $\sigma_F$ and $T$ \cite{claassen_position-momentum_2015, roy_band_2014, ledwith_fractional_2020-1}. In our study, we introduce a small sublattice potential (10$^{-7}$ eV) to isolate a Chern band, focusing on the conduction band due to particle-hole symmetry (the lower band after adding the small mass term). In the chiral limit, the trace condition is perfectly met at the magic angle, attributed to the holomorphic nature of the wave function \cite{tarnopolsky_origin_2019}, as shown in Fig. \ref{Fig3}(c). Deviations from the magic angle reveal a quadratic relationship between $T$ and the twist angle. Around the magic angle, $T$ is inversely proportional to the strength of Kane-Mele SOC, indicating that stronger Kane-Mele SOC enhances the conditions for achieving an FCI. Additionally, Fig. \ref{Fig3}(a) shows that finite Kane-Mele SOC smooths the relationship between $\sigma_F$ and the twist angle.

Moving away from the chiral limit, as illustrated by comparing Fig. \ref{Fig3}(c) and Fig. \ref{Fig3}(d), we observe that the trace condition is no longer perfectly met, though a quadratic relationship still holds. Similarly, Berry curvature deviations exhibit a more complex relationship with the twist angle. Despite larger magnitudes of both Berry curvature deviation and trace condition violation, they remain comparable to the critical values observed experimentally where FCI states have been detected \cite{xie_fractional_2021}.

To better understand the impact of AA hopping on the FCI, we conduct band-projected exact diagonalization (ED) calculations. These calculations assume full valley and spin polarization, focusing solely on the lowest conduction band. We use a long-range Coulomb interaction ($\frac{e^2}{\epsilon r}$) with a filling factor of $\nu = 1/3$ and $\epsilon = 5$. The results, presented in Fig. \ref{Fig3}(e)-(h), include a representative FCI spectrum shown in Fig. \ref{Fig3}(g). This spectrum features three quasi-degenerate states at momentum K=0, distinctly separated from higher excited states. Additionally, the particle entanglement spectrum, as shown in Fig. \ref{Fig3}(h), reveals 1710 states below the gap, corresponding to the quasi-hole excitations expected in the fractional quantum Hall effect.

The momenta, degeneracy, and particle-cut entanglement spectrum confirm the presence of FCI states \cite{regnault2011fractional, tang2011high, sun2011nearly, neupert2011fractional, sheng2011fractional}. The exact diagonalization (ED) results in Fig. \ref{Fig3}(e) show that the energy gap increases with Kane-Mele SOC, reaching a peak at a certain value. However, when AA hopping is included, the FCI state disappears for $\lambda_{KM} > 2$ meV, analogous to observations in MATBG. Despite this, a substantial parameter space remains where the FCI state is still stabilized. In summary, our ED results demonstrate that Kane-Mele SOC is crucial for stabilizing the FCI state in twisted bilayer graphene.

{\it Magic angle and flat bands in twisted Pt$_2$HgSe$_3$.---} 
 \begin{figure}
\includegraphics[width=0.5\textwidth]{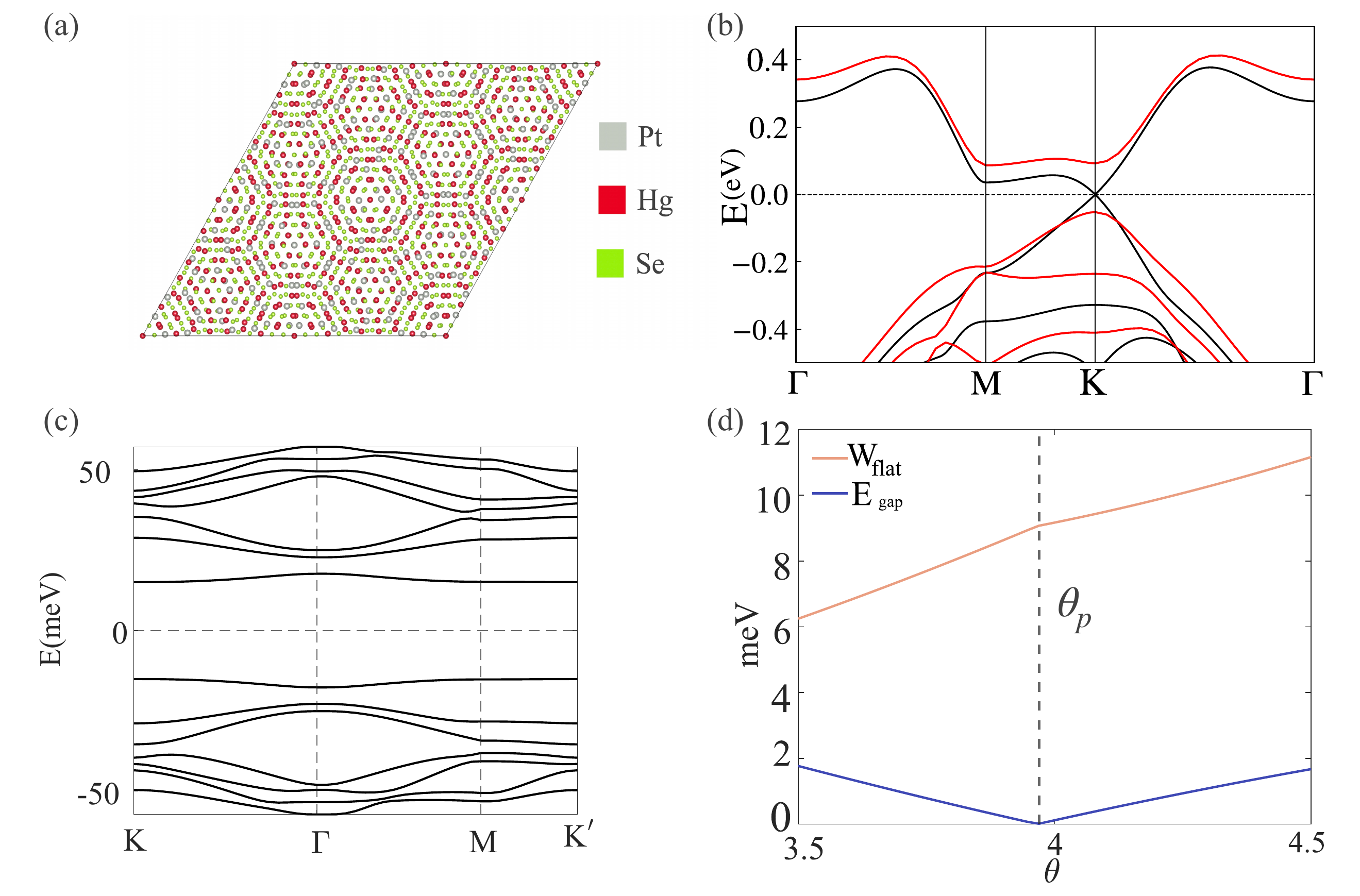}
     \caption{(a) The top view of twisted bilayer Pt$_2$HgSe$_3$. (b) The DFT band structure of monolayer Pt$_2$HgSe$_3$ with (red line)/without (black line) SOC. (c) The band structure of twisted bilayer Pt$_2$HgSe$_3$ at 2.7$^\circ$ with $w_0=0.7w_1$ from continuum model. (d) $W_{flat}$ denotes the band width of lowest conduction band and $E_{gap}$ represents the global gap between lowest moir\'e conduction band and remote moir\'e band. It is noteworthy that there are a gap close near 4.0 degree, which induces a topological phase transition.}
     \label{Fig4}
\end{figure}

In bilayer graphene with adjacent TMDs, the Kane-Mele SOC is three orders of magnitude smaller than interlayer tunnelling. However, for experimental realization, a system with large intrinsic Kane-Mele spin-orbit coupling is preferable. Therefore, we now study twisted bilayer Pt$_2$HgSe$_3$ as an alternative platform. Monolayer Pt$_2$HgSe$_3$ \cite{marrazzo_emergent_2020,marrazzo2018prediction} is predicted to be a quantum spin Hall insulator, with its low-energy physics well-described by the Kane-Mele model. Using parameters from \cite{rademaker_gate-tunable_2021, marrazzo2018prediction}, derived from tight-binding calculations, we model the bands of hBN-encapsulated twisted bilayer Pt$_2$HgSe$_3$ with a continuum model, as shown in Fig. \ref{Fig4}(c). We find that the magic angle for this system is approximately 2.7$^\circ$, and there is a pair of flat bands near the Fermi energy. However, unlike the topological flat bands in MATBG, these flat bands are trivial, as indicated by the Wilson loop spectrum in supplementary material. We will explain this in terms of the pseudo Landau level picture below.

It is well established that the topology of the flat bands in twisted bilayer graphene originates from the lowest pseudo-Landau level \cite{liu_pseudo-landau-level_2019, song_twisted_2021}. These lowest Landau levels are characterized by opposite sublattice polarizations, preventing coupling between them via AA stacking, and thereby preserving their topological nature. In contrast, higher Landau levels lack complete sublattice polarization, leading to significant coupling between them and a consequent loss of topological character. In the presence of Kane-Mele SOC, the lowest Landau level (LLL) is split into two components, one above and one below the Fermi level. When the Kane-Mele SOC strength is comparable to the gap between the LLL and the first Landau level (1stLL), the LLL can strongly hybridize with the 1stLL, leading to a trivial flat band. This situation is observed in twisted bilayer Pt$_2$HgSe$_3$. However, an intriguing phenomenon occurs as the twist angle increases. The gap between the LLL and the 1stLL, given by $\hbar \omega_c \propto \sqrt{w_1 \theta}$, where $\omega_c$ is the cyclotron frequency and $w_1$ is a characteristic parameter, enlarges with increasing twist angle. When this gap becomes sufficiently large, the LLL ceases to couple strongly with the 1stLL, enabling the emergence of a topological flat band. Numerical calculations, as depicted in Fig. \ref{Fig4}(d), reveal a topological phase transition around 4$^\circ$, indicating that a topological flat band is present when the twist angle exceeds 4$^\circ$.

{\it Conclusion.---}   
In this study, we focus on the twisted Kane-Mele model, which exhibits a relatively stable angle dependent electronic structures similar to twisted transition metal dichalcogenides (TMDs) \cite{zhang2021electronic}. In contrast, in magic-angle twisted bilayer graphene (MATBG), the relationship between bandwidth and twist angle is linear, so even a small change in twist angle significantly increases the bandwidth. The inclusion of Kane-Mele spin-orbit coupling (SOC) induces a quadratic relationship between bandwidth and twist angle. This leads to a stable van Hove singularity that remains largely unchanged with variations in the twist angle.

We find that stabilizing the magic angle helps correlated states resist angle disorder in practical devices. Specifically, by calculating the FCI indicator near the magic angle for different Kane-Mele SOC strengths, we show that the inclusion of Kane-Mele SOC leads to the quantum geometry favorable for FCI states, as confirmed by exact diagonalization. Our approach can be applied to other Kane-Mele model systems. For example, in twisted bilayer Pt$_2$HgSe$_3$, we find a magic angle of about 2.7° and a topological phase transition around 4°, suggesting a promising platform for future investigation of magic angle physics.

\section*{Acknowledgments}
We are grateful to Liang Fu, Andrei Bernevig and Amir Yacoby for helpful discussions. C.X. and Y. Z. are supported by the start-up fund at University of Tennessee Knoxville. The large matrix diagonalizations are performed on H100 nodes provided by AI Tennessee Initiative.


\bibliography{ref_my1.bib}
  
\newpage
\appendix
\clearpage

\onecolumngrid

\section{Tight-binding Calculation of MATBG with Kane-Mele SOC}

In this section, we calculate the band structure of magic-angle twisted bilayer graphene (MATBG) with the inclusion of Kane-Mele spin-orbit coupling (SOC), serving as a verification of the results presented in the main part. The tight-binding Hamiltonian we employ is given by:
\begin{equation}
	\begin{aligned}
		H=t\sum_{l,\langle i,j\rangle}(\boldsymbol{\hat{c}}^\dagger_{liA}\boldsymbol{\hat{c}}_{l jB}+H.c)+\mathrm{i}t_{KM}\sum_{l,s,s^\prime\langle\langle i,j \rangle\rangle}v_{ij}\boldsymbol{\hat{c}}^\dagger_{li\alpha}s^z_{s,s^\prime}\boldsymbol{\hat{c}}_{l j\alpha}+t_{\perp}(\boldsymbol{d})\sum_{i,j,\alpha,\beta}(\boldsymbol{\hat{c}}^\dagger_{1i\alpha}\boldsymbol{\hat{c}}_{2j\beta}+H.c)
	\end{aligned}
\end{equation}
with t stands for the amplitude of the intra-layer nearest-neighbour hopping, $t_{KM}$ is the strength of Kane-Mele SOC and $t_\perp$ is the inter-layer tunneling. And $l,s,i,\alpha$ is the index of layer, spin, unit cell and sub-lattice. $v_{ij}=\boldsymbol{\frac{r_i\times r_j}{|r_i\times r_j|}}=\pm 1$ depend on the hopping direction. Additionally, we apply the Slater-Koster approximation on inter-layer hopping:
\begin{equation}
\begin{aligned}
t_{\perp}(\boldsymbol{d})=V_{\sigma}^0\exp({-\frac{d-d_0}{l_0}})\big(\frac{\mathbf{d}\cdot \hat{\mathbf{e_z}}}{d}\big)^2
\end{aligned}
\end{equation}
where $d_0$ is the inter-layer distance ,  $\boldsymbol{d}$ is the position vector between the hopping sites and $l_0$ is the typical tunneling distance, we choose it to make the next-neighbour interlayer hopping be $0.1V_{\sigma}^0$ as usual. 
where  $d_0$  is the interlayer distance, $ \boldsymbol{d} $ is the position vector between the hopping sites, and  $l_0$  is the characteristic tunneling distance. We choose  $l_0$  to ensure that the next-nearest neighbor interlayer hopping is set to  $0.1V_{\sigma}^0$ , as is commonly done \cite{moon2013optical}.
The parameter we use:
\begin{equation}
\begin{aligned}
t & =-2.7 \ eV\\
V_{\sigma}^0&=0.48eV\\
d_0&=0.335nm\\
l_0 &=0.045nm
\end{aligned}
\end{equation}
\begin{figure}[h]
\centering
	\includegraphics[width=0.6\hsize]{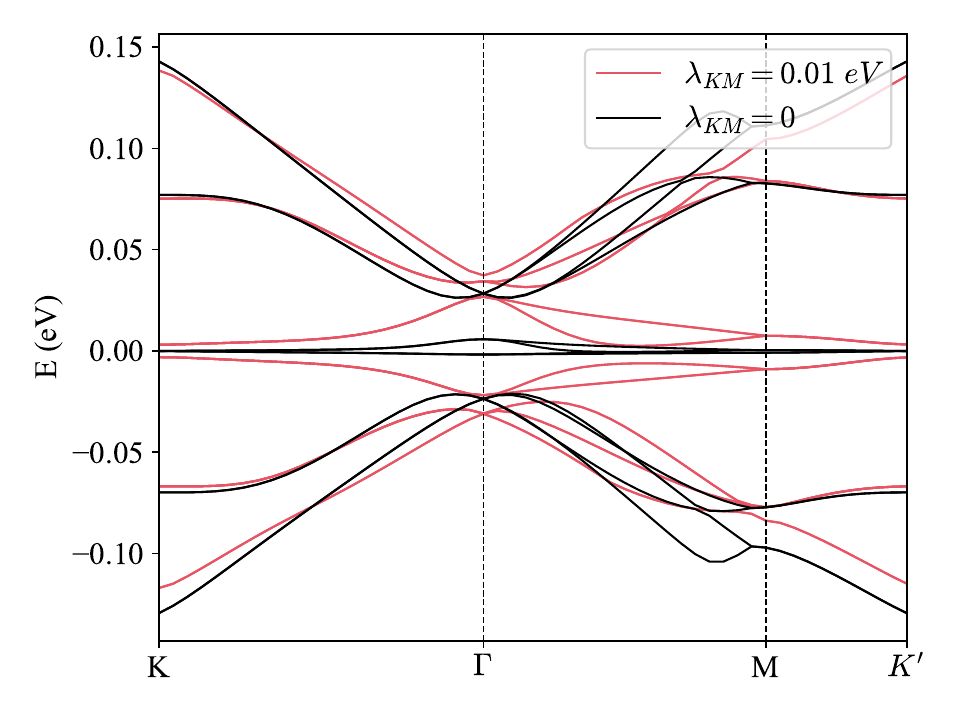}
	\caption{The band structure of twisted bilayer graphene with(red line)/without(black line) Kane-Mele SOC by the tight-binding model at angle $(m,n)=(31,30)$.($\lambda_{KM}=3\sqrt{3}t_{KM}$ here)}
\end{figure}

 We model the lattice deformation by:
\begin{equation}
	\begin{aligned}
&\boldsymbol{d}_{\boldsymbol{z} \mathbf{1}}(\boldsymbol{r})=\frac{d_0}{2}+d_1 \sum_{i=1}^3 \cos \left(\boldsymbol{G}_i^m \cdot \boldsymbol{r}\right) \\
&\boldsymbol{d}_{\boldsymbol{z} \mathbf{2}}(\boldsymbol{r})=-\frac{d_0}{2}-d_1 \sum_{i=1}^3 \cos \left(\boldsymbol{G}_i^m \cdot \boldsymbol{r}\right)
\end{aligned}
\end{equation}
with:
\begin{equation}
	\begin{aligned}
		d_0&=\frac{1}{3}(d_{AA}+2d_{AB})\\
		d_1&=\frac{1}{9}(d_{AA}-d_{AB})
	\end{aligned}
\end{equation}
where $d_{AA},d_{AB}$ is the distance between the AA,AB stacking region in twisted bilayer graphene and $\boldsymbol{G_i^m}$ is the reciprocal vector of moire unit cell. We choose the parameter to reproduce the distance from the reference\cite{liu_pseudo-landau-level_2019}:
\begin{equation}
	\begin{aligned}
		d_{AA}&\approx 3.60  \ \AA\\
		d_{AB}&\approx 3.35  \ \AA\\
	\end{aligned}
\end{equation}
Thus :
\begin{equation}
	\begin{aligned}
		d_0&=3.433 \ \AA\\
		d_1&=0.0278 \ \AA
	\end{aligned}
\end{equation}

\section{The bandwidth when Kane-Mele SOC is included }
\subsection{K points}
We make the Hamiltonian dimensionless by :
\begin{equation}
	\begin{aligned}
		E \rightarrow \frac{E}{\hbar v_F k_{\theta} } \ \textbf{and} \ \  k\rightarrow \frac{k}{k_\theta}
	\end{aligned}
\end{equation}
For simplify and feasibility, we choose the four plane wave $K-center \ model$ \cite{bernevig_twisted_2021}(we focus on the spin up here):
\begin{equation}
	\mathcal{H}_{k}=\left(\begin{array}{cccc}
h_{k} & T_{1} & T_{2} & T_{3} \\
T_{1}^{\dagger} & h_{k+q_{1}} & 0 & 0 \\
T_{2}^{\dagger} & 0 & h_{k+q_{2}} & 0 \\
T_{3}^{\dagger} & 0 & 0 & h_{k+q_{3}}
\end{array}\right)
\end{equation}
Solve the Schrodinger Equation:
\begin{equation}
	\mathcal{H}_k\psi_{k}=E_k\psi_k
\end{equation}
the wave function $\psi_k$ is a eight component vector
\begin{equation}
\psi_k=[\psi_0\ \psi_1\ \psi_2 \ \psi_3]	^{T}
\end{equation}
Write it in detail:
\begin{equation}
	\begin{aligned}
		&h_{k}\psi_0+\sum_{i=1}^3 T_i\psi_i=E_k\psi_0\\
		&T_1^{\dagger}\psi_0+h_{k+q_1}\psi_1=E_k\psi_1\Rightarrow \psi_1=(E_k-h_{k+q_1})^{-1}T_1^{\dagger}\psi_0\\
		&T_2^{\dagger}\psi_0+h_{k+q_2}\psi_1=E_k\psi_2\Rightarrow \psi_2=(E_k-h_{k+q_2})^{-1}T_2^{\dagger}\psi_0\\
	    &T_3^{\dagger}\psi_0+h_{k+q_3}\psi_1=E_k\psi_3\Rightarrow \psi_3=(E_k-h_{k+q_3})^{-1}T_3^{\dagger}\psi_0\\
	\end{aligned}
\end{equation}
Use the $Gaussian\  Elimination$:
\begin{equation}
\begin{aligned}
	&h_{k}\psi_0+\sum_{i=1}^3 T_i(E_k-h_{k+q_i})^{-1}T_i^{\dagger}\psi_0=E_k\psi_0\\
\end{aligned}
\end{equation}
We will apply the $Zero \ Angle$ approximation, because the 1st magic angle is so small.In this approximation, the onsite energy in k space reads:
\begin{equation}
	h_{k+q_i}=(k-q_i)\cdot \sigma+\lambda_{KM}\sigma_z
\end{equation}
For simplify, denote:
\begin{equation}
	\begin{aligned}
		& h_k=k\cdot \sigma+\lambda_{KM}\sigma_z
\\
		& h_{q_i}=-q_i\cdot \sigma
	\end{aligned}
\end{equation}
We focus on the flat band, whose energy is small compared to the remote bands, allowing us to treat $E_k$ as a perturbation
:
\begin{equation}
	\begin{aligned}
		(E_k-h_{k+q})^{-1}=	(E_k-h_{k}-h_q)^{-1} \approx -h_q^{-1}-h_q^{-1}(E_k-h_k)h_q^{-1}
	\end{aligned}
\end{equation}
Substituting it into the above equation:
\begin{equation}
\begin{aligned}
	&h_{k}\psi_0-\sum_{i=1}^3 T_i[h_{q_i}^{-1}+h_{q_i}^{-1}(E_k-h_k)h_{q_i}^{-1}]T_i^{\dagger}\psi_0=E_k\psi_0\\
	\Rightarrow & h_{k}\psi_0-\sum_{i=1}^3 \bigg[T_ih_{q_i}^{-1}T_i^\dagger+T_ih_{q_i}^{-1}(E_k-k\cdot \sigma-\lambda_{KM}\sigma_z)h_{q_i}^{-1}]T_i^{\dagger}\bigg ]\psi_0=E_k\psi_0\\
	\Rightarrow &( k\cdot \sigma+\lambda_{KM}\sigma_z)\psi_0-3(w_0^2+w_1^2)E_k\psi_0-3w_1^2k\cdot \sigma\psi_0+3\lambda_{KM}(w_1^2-w_0^2)\sigma_z\psi_0=E_k\psi_0\\
	\Rightarrow & \bigg[ E_k-\frac{1-3w_1^2}{1+3w_0^2+3w_1^2}k\cdot \sigma-\frac{1+3(w_1^2-w_0^2)}{1+3w_0^2+3w_1^2}\lambda_{KM}\sigma_z \bigg]\psi_0=0
\end{aligned}
\end{equation}
Thus the energy spectrum reads:
\begin{equation}
	E_k=\sqrt{\bigg(\frac{1-3w_1^2}{1+3w_0^2+3w_1^2}\bigg)^2|\boldsymbol{k}|^2+\bigg(\frac{1+3(w_1^2-w_0^2)}{1+3w_0^2+3w_1^2}\bigg)^2\lambda_{KM}^2}
\end{equation}
For convenience, denote:
\begin{equation}
	\begin{aligned}
		A&=\frac{1-3w_1^2}{1+3w_0^2+3w_1^2}\\
		B&=\frac{1+3(w_1^2-w_0^2)}{1+3w_0^2+3w_1^2}
	\end{aligned}
\end{equation}
Owing to $|\boldsymbol{k}|$ is a small parameter, we can expand the spectrum near the $K$:
\begin{equation}
	E_k=\lambda_{KM}B+\frac{A^2}{2\lambda_{KM}B}|\boldsymbol{k}|^2
\end{equation}
The linear dispersion is transformed into the parabolic dispersion.
\subsection{$\Gamma$ points}
For simplify and feasibility, we choose the six plane wave $\Gamma-center \ model$:
\begin{equation}
	\mathcal{H}_{k}=\left(\begin{array}{cccccc}
h_{k-q_1} & T_{2} & 0 & 0 & 0 & T_{3} \\
T_{2} & h_{k+q_3} & T_{1} & 0 & 0 & 0 \\
0 & T_{1}& h_{k-q_2}& T_{3}& 0 & 0 \\
0 & 0 & T_{3} & h_{k+q_1}& T_{2} & 0 \\
0 & 0 & 0 & T_{2} & h_{k-q_3} & T_{1}\\
T_{3}& 0 & 0 & 0 & T_{1}& h_{k+q_2}
\end{array}\right)
\end{equation}
Solve the Schrodinger Equation:
\begin{equation}
	\mathcal{H}_k\psi_{k}=E_k\psi_k
\end{equation}
we can solve it at $k=0$, the energy of the lowest conduction bands reads:
\begin{equation}
	E_{\Gamma}=\sqrt{1+\lambda_{KM}^{2}+w_{0}^{2}+4 w_{1}^{2}-2 \sqrt{\lambda_{KM}^{2} w_{0}^{2}+4 w_{1}^{2}+4 w_{0}^{2} w_{1}^{2}}}
\end{equation}
Based on the solution at $K$, we have:
\begin{equation}
	E_{K}=\frac{1+3w_1^2-3w_0^2}{1+3w_0^2+3w_1^2}m
\end{equation}
Thus the bandwidth of the flat band:
\begin{equation}
	W=E_{\Gamma}-E_{K}
\end{equation}
To study the angle-dependent behavior, we need to reintroduce the dimensional energy, and we choose:
\begin{equation}
	\begin{aligned}
&		w_1=110\ meV \\
&       w_0=\alpha \  w_1\\
&       \hbar v_F=5.944 eV\cdot \AA \\
&       |K|=\frac{4\pi}{3a}=1.7028\AA^{-1} \\
&       E_{\theta}=\hbar v_F|K|\theta \\
	\end{aligned}
\end{equation}
In chiral limit($w_0=0$):
\begin{equation}
	\begin{aligned}
		W&=\sqrt{1+\lambda_{KM}^{2}+4 w_{1}^{2}-4 w_1}-\lambda_{KM}\\
		&=\sqrt{\lambda_{KM}^{2}+(1-2w_1)^2}-\lambda_{KM}\\
		&=\sqrt{\lambda_{KM}^{2}+(E_{\theta}-2w_1)^2}-\lambda_{KM}\\
		&=\sqrt{\lambda_{KM}^{2}+(\hbar v_F|K|\theta-2w_1)^2}-\lambda_{KM}
\end{aligned}
\end{equation}
Thus the minimum of the bandwidth in chiral limit:
\begin{equation}
	W=0
\end{equation}
The magic angle defined by the bandwidth is 0.5, which is smaller than the exact value of 0.587 obtained from numerical results or perturbation theory \cite{tarnopolsky_origin_2019}. This discrepancy arises because only six plane waves are considered here. While these approximations do not perform well in accurately evaluating the magic angle, they provide a qualitative description of the bandwidth. Notably, the relationship between the bandwidth and the twist angle is quadratic.
\begin{equation}
\begin{aligned}
&\frac{	\partial W}{\partial \theta}=\frac{|K| v_F \left(|K| v_F \theta-2 w_1\right)}{\sqrt{\left(|K| v_F \theta-2 w_1\right)^2+\lambda_{KM}^2}}=0	\\
&\frac{	\partial^2 W}{\partial \theta^2}=\frac{|K|^{2} v_ F^{2}}{\sqrt{\lambda_{KM}^{2}+\left(|K| v_F \theta-2 w_{1}\right)^{2}}}-\frac{|K|^{2} v_F^{2}\left(|K| v_F \theta-2 w_{1}\right)^{2}}{\left(\lambda_{KM}^{2}+\left(|K| v_F \theta-2 w_{1}\right)^{2}\right)^{3 / 2}} \bigg|_{|K| v_F \theta=2 w_1}                              \\
&\ \ \ \ \ \ \ =\frac{\hbar^2|K|^2v_F^2}{\lambda_{KM}}
\end{aligned}
\end{equation}

\section{ The comparison between Kane-Mele SOC and sub-lattice potential.} 
Previous experiments \cite{xie_fractional_2021, serlin2020intrinsic, sharpe2021evidence} have investigated hBN-aligned MATBG, which introduces a sublattice potential on the adjacent layer, creating a gap between the flat bands and allowing them to carry a Chern number. In fact, hBN-aligned MATBG exhibits the same band dispersion pattern, as the system with Kane-Mele spin-orbit coupling discussed earlier. Below, we analyze this in the context of the pseudo Landau level representation of the flat band, which is spanned by the pseudo-Landau levels $\psi_{A1}$ and $\psi_{B2}$. In the chiral limit, the Hamiltonian for hBN-encapsulated MATBG can be approximated as follows:
\begin{equation}
	H_m=m
	\begin{bmatrix}
		1 & 0 \\
		0 & -1 \\
	\end{bmatrix}
\end{equation}
which is the same as the Kane-Mele SOC that works equally on both layers, thus they share the same band dispersion. Therefore, the conclusions drawn from analyzing the Kane-Mele spin-orbit coupling also apply when considering the sub-lattice potential. Both the Kane-Mele SOC and the sub-lattice potential contribute to the stabilization of the electron structure in magic-angle twisted bilayer graphene.

However, there are crucial differences in underlying wavefunctions. The Chern number of the flat band is determined by the direction of the pseudo-magnetic field, causing the lowest Landau level to be fully polarized on different sub-lattices. Therefore, we can use the sub-lattice index to label the Chern number (e.g., A means C=1). When the Kane-Mele spin-orbit coupling is included, the components of the conduction band are $\psi_{A1\uparrow}$ and $\psi_{B2\downarrow}$, which carry opposite Chern numbers. Consequently, this model can serves as an excellent platform for studying the fractional quantum spin Hall effect \cite{wu2024time}. In contrast, the sub-lattice potential does not distinguish between spins; thus, the conduction band consists of $\psi_{A1\uparrow}$ and $\psi_{A1\downarrow}$, both carrying the same Chern number. Moreover, the sublattice potential induced by hBN is approximately 30 meV \cite{zhang2019twisted}, which interferes with the conditions required for the formation of a fractional Chern insulator (FCI). This disruption likely contributes to the absence of FCI detection at zero magnetic field in magic-angle twisted bilayer graphene.

\section{The topological phase transition in twisted bilayer Pt$_2$HgSe$_3$}
The monolayer Pt$_2$HgSe$_3$ is predicted to be a quantum spin hall insulator with large gap and its low energy can be described by the Kane-Mele model. We describe the twisted bilayer Pt$_2$HgSe$_3$ with the continuum model:
\begin{equation}
\begin{aligned}
	H&=	\hbar v_F\boldsymbol{k\cdot \sigma}\tau_0s_0+\lambda_{KM}\sigma_zs_z\tau_0+m\sigma_z\tau_0s_0+
	\begin{bmatrix}
	0 & \boldsymbol{\hat{T}{(r)}}\\	
	\boldsymbol{\hat{T}^{\dagger}{(r)}} & 0\\
	\end{bmatrix}\otimes s_0
\end{aligned}
\end{equation}
we use the parameter\cite{rademaker_gate-tunable_2021, marrazzo2018prediction} :
\begin{equation}
	\begin{aligned}
		\hbar v_F&=\frac{\sqrt{3}at}{2}=1.2032 \ eV\cdot \AA\\
		\lambda_{KM}&=3\sqrt{3}t_{KM}=107.56\ meV\\
		w_1&=18 \ meV \\
		m & = 135 meV \\
	\end{aligned}
\end{equation}
\begin{figure}
	\includegraphics[scale=0.15]{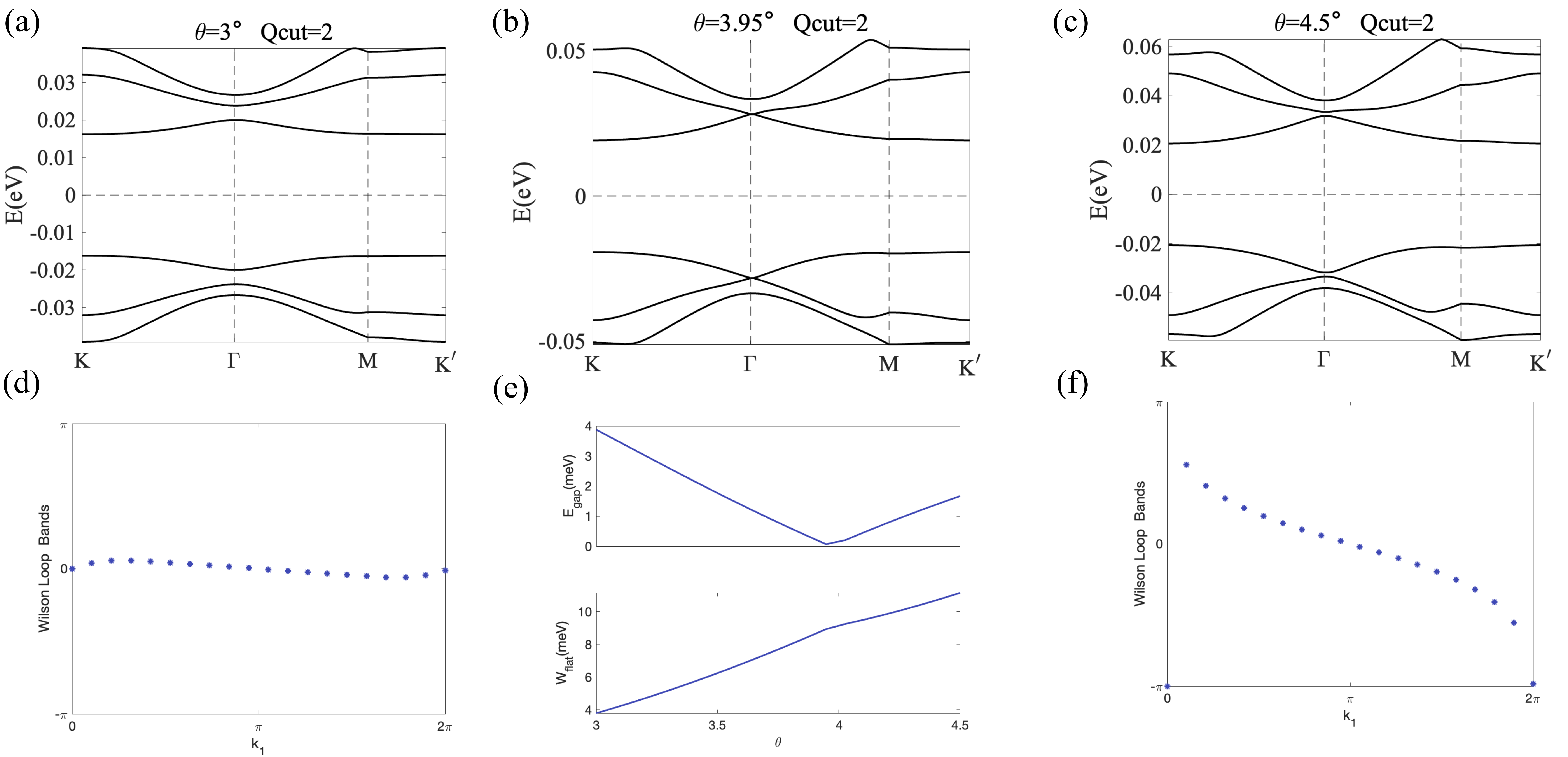}
	\caption{(a)(b)(c) is the band structure of twisted bilayer Pt$_2$HgSe$_3$ at different angle we choose $w_0=0.7w_1$ here.(d)/(f) is Wilson loop bands of lowest conduction band at 3$^\circ$/4.5$^\circ$. (e) shows the band width of lowest conduction band and the gap between the flat  band and remote band.}
 \label{Fig_SM1}
\end{figure}

There are a topological phase transition in this system as we shown in Fig. \ref{Fig_SM1}. we also give a picture to understand it here: If we introduce the Kane-Mele spin-orbit coupling (SOC) or a  sub-lattice potential equally across both layers, the energy of the Landau levels will shift without altering the wave function. When this effect is small, i.e., $\lambda_{KM}\ll \hbar w_c$ , the Landau levels remain largely unaffected by the remote bands, preserving their topological properties. However, if we continue to increase the strength of the Kane-Mele SOC until it becomes comparable to the energy gap, the zeroth Landau level will strongly couple with the first Landau level in another sub-space, leading to the loss of their topological characteristics. This indicates the occurrence of a topological phase transition.
The typical band gap between the flat band and the remote band is :
\begin{equation}
	\hbar w_c=\sqrt{\frac{8\pi\hbar v_F w_{1}\theta}{a}} \propto \sqrt{w_{1}\theta}
\end{equation} 

The topological phase transition will happen when:
\begin{equation}
\begin{aligned}
	\lambda_{KM}\sim \sqrt{w_{AB}\theta}\ \ &\Rightarrow \ \frac{\lambda_{KM}^2}{\theta} \sim w_{AB}\\
	\frac{\lambda_{KM}^2}{\theta} \gg w_{AB}&:\ \ \text{Trivial} \\
		\frac{\lambda_{KM}^2}{\theta} \ll w_{AB}&:\ \ \text{Topological}
\end{aligned}
\end{equation}
Thus: for a specific value of the Kane-Mele SOC, a small twist angle corresponds to a trivial phase, while a larger twist angle induces a topological phase. When AA hopping is introduced, the topological phase transition will happen when:
\begin{equation}
	\lambda_{KM}+C\frac{w_{AA}^3}{(\hbar w_c)^2}\sim \hbar w_c
\end{equation}
Where C is the constant. This suggests that AA hopping will cause the phase transition to occur at a smaller Kane-Mele SOC strength and at a larger twist angle. When both the Kane-Mele SOC and mass term are applied to this system, there are some differences. However, they can still be explained within the same framework.

\section{The FCI indicator}
The band geometry is encoded in the Quantum Geometry Tensor: 	
\begin{equation}
		\eta^{uv}:= A_{BZ}\langle \partial^uu_{\boldsymbol{k}}\mid (1-|u_{\boldsymbol{k}}\rangle \langle u_{\boldsymbol{k}} |)\mid \partial^vu_{\boldsymbol{k}}\rangle
\end{equation}
where $u ,v$ is the label of the Cartesian coordinate and $A_{BZ}$ is the area of the Brillouin zone, it makes this tensor dimensionless. Its symmetric and antisymmetric part separately give the Berry curvature ($\Omega(\boldsymbol{k})=-2\mathrm{Im}(\eta^{xy})$) and quantum metric ($g^{uv}(\boldsymbol{k})= \mathrm{Re}(\eta^{uv})$). Here we introduce two parameters to describe the band geometry:
\begin{equation}
	\begin{aligned}
		\sigma_F :& =\bigg[\frac{1}{A_{BZ}} \int d^2{\boldsymbol{k}}(\frac{\Omega(\boldsymbol{k})}{2\pi}-1)^2\bigg]^{\frac{1}{2}}\\
		T:&= \frac{1}{A_{BZ}} \int d^2{\boldsymbol{k}}\bigg[tr(g(\boldsymbol{k}))-\Omega(\boldsymbol{k})|\bigg]
	\end{aligned}
\end{equation}
The $\sigma_F $ describes the fluctuations of Berry curvature and the T indicates the violation of the trace condition .

\end{document}